\documentclass[runningheads,a4paper]{llncs}

\usepackage{graphicx}
\usepackage{epstopdf}
\usepackage{amssymb}
\usepackage{tabularx}
\usepackage{tabulary}
\usepackage{booktabs}
\usepackage{float}
\usepackage[hyphens]{url}
\usepackage{hyperref}
\usepackage[all]{hypcap}
\usepackage{tabto}
\usepackage{amsmath}
\usepackage{enumitem}
\usepackage[autostyle]{csquotes}
\usepackage{listings}
\usepackage{subfig}
\usepackage{color}

\usepackage[numbers,square]{natbib}

\newcommand{\keywords}[1]{\par\addvspace\baselineskip
\noindent\keywordname\enspace\ignorespaces#1}

\renewcommand\bibsection[1]{\section*{\refname}\small #1} 

\begin{document}

\mainmatter

\title{Archiving Software Surrogates on the Web\\for Future Reference\thanks{This work is partly funded by the German Research Council under FID Math and the European Research Council under ALEXANDRIA (ERC 339233)}}

\titlerunning{Software in Web Archives}

\author{Helge Holzmann\inst{1} \and Wolfram Sperber\inst{2} \and Mila Runnwerth\inst{3}}

\authorrunning{Holzmann et al.}

\institute{L3S Research Center\\
Appelstr. 9a, 30167 Hannover, Germany, \email{holzmann@L3S.de}\\
\and
zbMATH, FIZ Karlsruhe - Leibniz Institute for Information Infrastructure\\
Franklinstr. 11, 10587 Berlin, Germany, \email{wolfram@zentralblatt-math.org}\\
\and
German National Library of Science and Technology (TIB)\\
Welfengarten 1b, 30167 Hannover, Germany, \email{Mila.Runnwerth@tib.eu}}

\maketitle

\begin{abstract}
Software has long been established as an essential aspect of the scientific process in mathematics and other disciplines. However, reliably referencing software in scientific publications is still challenging for various reasons. A crucial factor is that software dynamics with temporal versions or states are difficult to capture over time. We propose to archive and reference surrogates instead, which can be found on the Web and reflect the actual software to a remarkable extent. Our study shows that about a half of the webpages of software are already archived with almost all of them including some kind of documentation.

\end{abstract}

\section{Introduction}
\label{sec:introduction}

Software is used in science among all disciplines, from analysis software and supporting tools in the humanities, over controlling and visualization software in medicine to the extensive use of all kinds of software in computer science as well as mathematics. However, referencing software in scientific publications has always been challenging. One reason is that software alone is often not considered a scientific contribution and therefore, properly citable publications do not exist. This is particularly an issue if the software does not tackle a concrete research question, but was created as tool for various purposes or different kinds of research, such as standard software like \textit{Microsoft Excel}. Another issue is, even if the software in question is well-known or even published, it undergoes dynamics and the version an article refers to might be different from the one currently available. Furthermore, a name like \textit{Microsoft Excel} refers to a \textbf{product} rather than a concrete version of that software. Also publications typically deal with the innovation and benefits created by the software as a product rather than a concrete build, version or setup. However, this very specific \textbf{artifact} may be crucial to reconstruct a software instance as in the original experimental setup, to reproduce experiments and comprehend scientifically published results.

As an example consider the famous bug of \textit{Excel 2007}, which produced the number \texttt{100,000} in a cell of which the underlying data equaled to \texttt{65,535}\footnote{\url{http://blog.wolfram.com/2007/09/25/arithmetic-is-hard-to-get-right} [from 25/09/2007]}. To fix this, the appropriate patch was released shortly after\footnote{\url{https://support.microsoft.com/en-us/kb/943075} [from 09/10/2007]}, which resulted in an artifact with an updated minor version number, but of course did not update the major version \textit{2007}. Actually, even though 2007 in this case is already more specific than just the product's name, it should be better considered a \textbf{sub-product} of the \textbf{product family} \textit{Excel} rather than a concrete version, since it does not refer to a concrete artifact. Therefore, to verify results of \texttt{100,000} in scientific experiments with \textit{Excel 2007} involved, the precise version number referring to the exact artifact used in the experiment is required, but very unlikely to be mentioned in a publication.

In the context of our project \textit{FID Math}, aiming for a mathematical information service infrastructure, we are facing the problem of referencing software with a focus on all kinds of mathematical applications, tools, as well as services. In the area of math, software is heavily used for various purposes, such as calculations, simulations, visualizations and more. Very often multiple are combined, while the critical task is performed by a script, which is software in itself, running inside an environment like \textsf{MATLAB}, \textsf{Mathematica} or \textsf{Sage}. Settings like these make it particularly challenging to reference a consistent state of the incorporated software. Further, the mix of open source and proprietary software introduces an additional challenge due to crucial differences in many aspects, such as code contributions, licensing, as well as the question for preservation. To address these difficulties in a universal manner, we propose Web archiving as a solution to preserve representations of software on the Web as surrogates for future reference.

\section{Problem and Questions}
\label{sec:problem}

In an ideal world, the results of every experiment conducted and published by scholars in their scientific work should be reproducible. This in turn implies that every software can be recovered in the exact state as used in their experiments. This either requires a detailed reference to the software's state and general access to software artifacts, or, alternatively, ways to freeze and preserve a software's state and provide it as attachment of a publication. Both seems unrealistic for practical as well as legal reasons. While open source projects often suffer a reliable release process with proper versioning, every committed state is usually precisely identified by a single hash, such as the \textit{SHA} used by \textsf{GIT}\footnote{\url{https://git-scm.com}}. This hash does not only encode the current state of the software, but refers to all previous commits comprising relevant meta data records. By contrast, structured meta data of proprietary software is often more explicit, with the author being the company behind the software and each bugfix or patch presumably increases the minor version number of the software. Accordingly, both types of software potentially allow referring to concrete artifacts. The challenge is to establish a unified representation and ways to recover the referenced software.

Open source licenses commonly allow redistribution, which facilitates sharing preservation copies with publications to replay experiments. For proprietary software this is usually considered piracy. Even more difficult to handle are Web APIs and services, where the user does not have access to the actual software, but only to the interface. However, by recovering in this context we do not necessarily mean to obtain a copy of the software, which is only required for replaying experiments and in many cases not usable without proper documentation anyway. Recovering can also mean to get an understanding of the software, for example through its \textit{documentation}, \textit{source code}, \textit{related publications} or \textit{change logs}. Already a brief description can be difficult to obtain though, as referenced software, after many years, might not even exist anymore.

Since the Web can be considered our primary source of all kinds of information today, our hypothesis is that most of the information listed above is available on the Web as well. Therefore, a snapshot of the corresponding websites of a software from when it was acquired for scholarly use, whether by downloading a copy or ordering in a shop, would constitute a representation of the software at that time. Although it might not include the artifact itself, it is the most comprehensive representation we can get, given the practical and legal restrictions. Hence, it can be considered a temporal surrogate of the actual software.

To realize a Web archiving solution for such a purpose, which enables reliably referencing software surrogates on the Web in scientific publications, we need to overcome a number of challenges. The system has to be aware of all relevant resources of a software and it has to ensure that these resources are preserved at the time of scholarly use. This will only work with manual intervention. Therefore, a concrete implementation idea remains for future work and is out of the scope of this paper. However, with existing Web archives we can already create a solution based on the publication dates of articles using software. In this study we ask the following questions to analyze the applicability of archiving software surrogates on the Web:

\begin{enumerate}[align=left, noitemsep, nolistsep]
\item[(\textbf{Q1})] How well is software represented by its surrogate on the Web?
\item[(\textbf{Q2})] Which information of software is available on the Web?
\item[(\textbf{Q3})] How many websites of mathematical software are archived?
\item[(\textbf{Q4})] For how many of these can referenced versions from the past be recovered?
\end{enumerate}

\section{Related Work}

The presented study touches different areas of interest from \textit{scientific software management}, dealing with the scientific process and software citations, over the question of how to preserve artifacts in \textit{software repositories}, to \textit{Web archives} and their role in science.

\subsection{Scientific Software Management}

Since software has become an essential part of scientific work in the field of mathematics but also other disciplines, various initiatives for research data management have begun to focus on this aspect as well. \citet{pengrepr} addresses the need for reproducibility of computational research. In 2011 research was unthinkable without software-based experiments, its findings had no platform to be published properly. Hence, the claim of replicating experiments was technically possible but had no place in the publishing process. In 2012, \citet{wilsbest} published a study on scientific software. It discovered that the majority of scientists use software for their research. It offers best practices to software development for research purposes because most scientists have never been taught to do so.

The Software Sustainability Institute\footnote{\url{http://www.software.ac.uk/}} (SSI) is a British facility with the objective to improve the role of scientific software. It is associated with the the slogan \textit{``Better software, better research''} coined by \citet{goblbet}. The SSI discusses how software can be developed, archived and referenced in order to contribute to scientific knowledge. Closely associated is the Digital Curation Centre\footnote{\url{http://www.dcc.ac.uk/}} with a more general mandate towards research data\citep{rusbdigi}. Funded by the German Research Foundation, \textsf{SciForge}\footnote{\url{http://www.sciforge-project.org/}} presents a concept to accompany software throughout the whole research process in a transparent and scientifically adequate manner.

Scientific standards for the handling of mathematical software are mentioned by \citet{vogtsoft}. On the one hand, researchers must be aware of how to develop software effectively. On the other hand, publishing habits must be adapted to archiving, citing and quality-approved software. As an example, the journal of \textit{Mathematical Programming Computation} investigates software for its scientific impetus\footnote{\url{http://mpc.zib.de/}}. Running the software is part of their reviewing process. FAIRDOM\footnote{\url{http://fair-dom.org/}} is a research data management initiative in systems biology. It proclaims a \emph{fair} use of research data where \emph{fair} is an acronym for findable, accessible, interoperable and reproducible~\citep{stanevol}. A study with a focus on repeatability of computational research shows that standards are required to reconstruct published research on computational systems~\citep{collrepe} In 2012, Mike Jackson published two blog articles on how to cite and describe software\footnote{\url{http://software.ac.uk/so-exactly-what-software-did-you-use}}\footnote{\url{http://www.software.ac.uk/blog/2012-06-22-how-describe-software-you-used-your-research-top-ten-tips}}. Until today they influence the discussion about software citation. Psychology has adopted a citation standard in order to cite software in a standardized way~\citep{paizpub}. For the humanities, the Modern Language Association of America (MLA) provides a styleguide to cite software~\citep{gibamla}. Among many disciplines and especially in mathematics, \textit{BibTeX} and \textit{BibLaTeX} are widespread tools to manage bibliographies. Although they are adaptable, there is currently no approved standard to give a distinct reference to all varieties of software.

\subsection{Software Repositories}

There are various institutional and domain-specific repositories for research data. An overview is given by the Registry of Research Data Repositories\footnote{\url{http://www.re3data.org/}}~\citep{pampmaki}. Many of these repositories are operated by universities or have been initiated by research institutes. The University of Edinburgh offers a research data repository, DataShare~\citep{macdedin}. It is divided into domain-specific collections which can be searched separately or compositely. Harvard's DataVerse Project\footnote{\url{http://dataverse.org/}} supplies a web service to share, archive and cite research data. \textsf{RADAR}, funded by the German Research Foundation, pursues a service-oriented approach to provide infrastructure and services to host research data repositories~\citep{krafrada}. However, currently, there is no designated repository for mathematical software beyond institutional level.

\subsection{Web Archives}

Preservation of Web resources has recently been of growing interest, resulting in a considerable number of publications around Web archives from different perspectives. First, Web archives have been gaining growing popularity as scholarly source \citep{hockx2014access} in disciplines like the humanities \citep{gomes2014importance}. Second, Web archives have been of interest as subject of research themselves. In 2011, \citet{ainsworth2011much} analyzed how much of the web is archived and found that for up to 90\% of the pages in the considered collections at least one archived version exists, however, only a few of them have a consistent coverage over time. Other works in this context focused on a particular subset of the Web, such as national domains \citep{alkwai2015well, holzmann2016jcdl}, while our analysis has a very concentrated scope of mathematical software in Web archives. Beside the question of how much is archived, we also look at what resources are available, which falls in the area of profiling Web archive collections \citep{alsum2014profiling, day2014implementing, alam2015web}. Finally, the goal of this study is to investigate the applicability of Web archiving to preserve representative surrogates of software. Even though this particular subject has not been tackled before, other researchers looked into Web archiving to create preservation copies of other types or Web resources, such as blogs \citep{kasioumis2014towards} and social networks \citep{marshall2014argument}. \citet{salah2012losing} found a nearly linear relationship between time and the percentage of lost social media resources.

Another challenge that we are facing is to find the resources related to a particular software in a Web archive. One approach in this direction is to use secondary data sources with temporally tagged URLs, like the social bookmarking platform Delicious. This can be used for temporal archive search as demonstrated by the \textsf{Tempas} system \citep{holzmann2016www, holzmann2016sigir}.

\section{Data and Methodology}

The primary source for this work was \textsf{swMATH}, an information service for mathematical software\footnote{\url{http://www.swmath.org}} ((M)SW). Based on the information of SW in this directory, we analyzed the linked URLs on the current Web as well as in a Web archive, provided by the \textit{Internet Archive}\footnote{\url{http://archive.org}}.

\subsection{\textsf{swMATH} in a Nutshell}
\label{sec:swmath}

\begin{figure}[t]
	\centering
	\includegraphics[width=\textwidth]{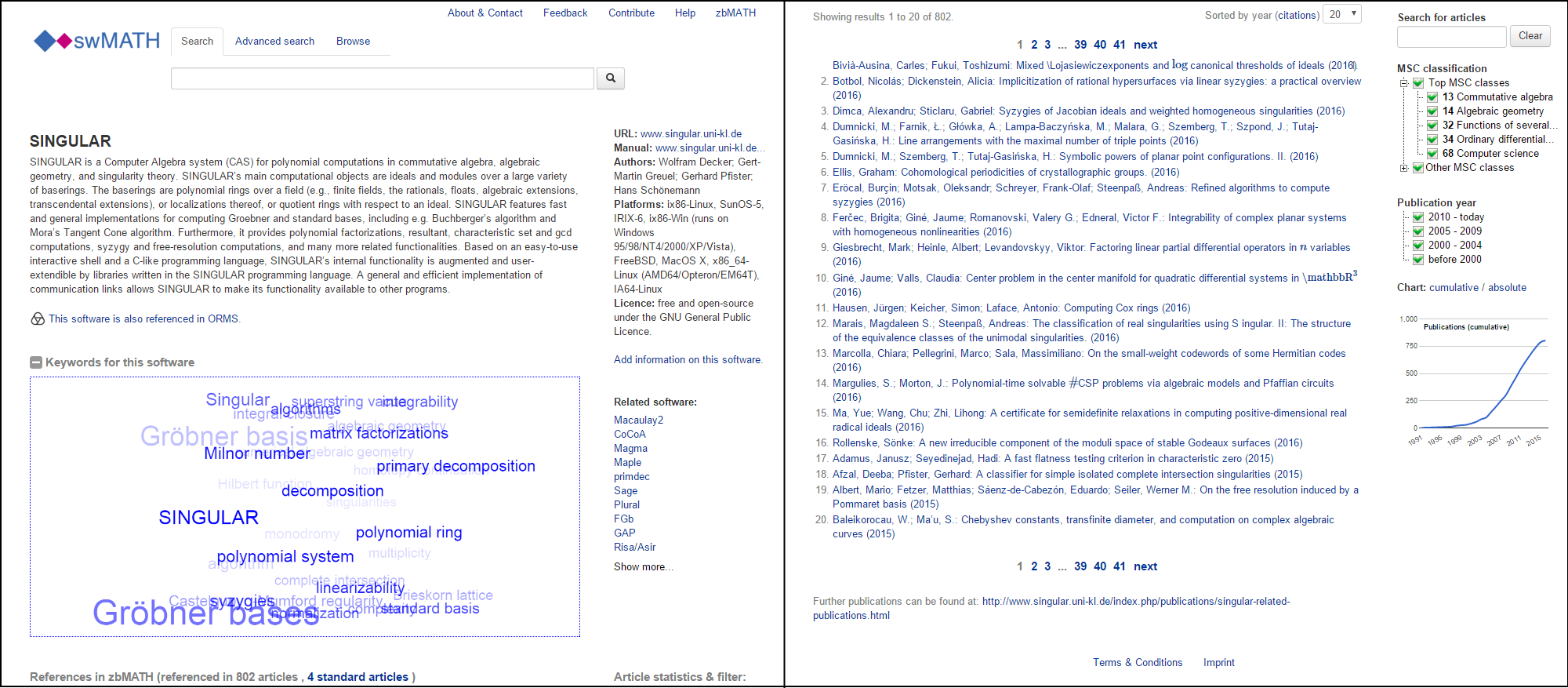}
	\caption{Mathematical Software \textsf{Singular} on \textsf{swMath}}
	\label{fig:swMath}
\end{figure}

\textsf{swMATH} is one of the most comprehensive information services for MSW \citep{greuel2014swmath}. It contains more than 12,000 records, each representing a SW \textbf{product} or \textbf{product-family} (cp. Sec.~\ref{sec:introduction}) with a unique identifier, as shown in Figure~\ref{fig:swMath}. \textsf{swMATH} is based on the database of \textsf{zbMATH}\footnote{\url{http://www.zbmath.org}}, one of the most comprehensive collections of mathematical publications, with more 110,000 articles referring to MSW. The biggest challenge for a service like \textsf{swMATH} is to recognize these references. In many cases, only a name is mentioned, while a version or an explicit label as (M)SW is missing. \textsf{swMATH} tackles this with simple heuristics, by scanning titles, abstracts, as well as references of publications to detect typical terms, such as \textit{solver}, \textit{program}, or simply \textit{software}, in combination with a name. 

After new candidates have been detected, they are checked manually to ensure the high quality of the service. As part of this manual intervention step, additional meta data, such as the \textbf{URL} of a SW is added. Later on, websites are periodically checked and outdated URLs are removed or replaced. In case there is no permanent link that points to a website of the SW, the URLs of a corresponding repository record or a publication is used instead.

Another important feature for our analysis is the \textbf{publication list} for every SW on \textsf{swMATH}. Each article in this list is annotated with its publication year. The publications can be sorted chronologically or by the number of citations an article has received. In \textsf{swMATH}, publications also serve as source for additional information, such as related software and the keyword cloud shown in every record (s. Fig.~\ref{fig:swMath}).

In order to enhance the functionality of \textsf{swMATH}, one goal is to capture the dynamics of (M)SW as reflected by the publications over time. We address this aspect by investigating which information of SW is available on the Web (s. Sec.~\ref{sec:software_on_the_web}) and can be recovered from Web archives (s. Sec.~\ref{sec:software_on_the_web}). Of importance for this study are \textbf{URLs} as well as the \textbf{publications} of a MSW, which are both available through \textsf{swMATH}.

\subsection{Analysis}
\label{sec:analysis}

As an initial step of each analysis in our study (s. Sec.~\ref{sec:results}), we crawled the required datasets using \textsf{Web2Warc}\footnote{\url{https://github.com/helgeho/Web2Warc} (\textit{Last commit 73f0934 on Jan 29, 2016})}, resulting in the following four collections, listed with the last time of crawl:

\begin{itemize}[align=left, noitemsep]
\item \textbf{swMATH records} \textit{(28/01/2016 - 14:14:53)}\\
All 11,785 software pages available on \textsf{swMATH} at that time.

\item \textbf{URLs} \textit{(18/02/2016 - 18:55:03)}\\
All webpages linked by the 11,125 URLs extracted from the \textsf{swMATH} records.

\item \textbf{Publications} \textit{(19/02/2016 - 08:30:15)}\\
The top 100 user publications with respect to their number of citations received, for all \textsf{swMATH} records. These lists are dynamically loaded into the \textsf{swMATH} records and constitute therefore separate resource to be crawled.

\item \textbf{Internet Archive} \textit{(22/02/2016 - 19:38:55)}\\
Meta data of the captures in the Internet Archive's Web archive for all URLs extracted from \textsf{swMATH}, using the \textit{Wayback CDX Server API}\footnote{\url{https://archive.org/help/wayback_api.php}}. For each URL we fetched the latest capture as well as the one closest to the time of the best publication of the corresponding software with respect to number of citations received.
\end{itemize}

To analyze these collections, which are small Web archives in themselves, we employed \textsf{ArchiveSpark}\footnote{\url{https://github.com/helgeho/ArchiveSpark} (\textit{Last commit acc5a16 on Feb 17, 2016})}, a framework for Web archive analysis \citep{archivespark}. This way, we extracted the data of interest, such as URLs and publication dates from the \textsf{swMATH} records as well as linked URLs from the crawled webpages. Subsequently, URLs were analyzed further to identify what kind of resources they point to. We split the URLs by means of the following rules: 1. the host is split at dots [.], 2. the path is split at [/.-\_], 3. the query string is split at [?=+\&:-\_]. As indicated by the segments we obtained through those splits, the URLs were classified as different resource type. The classes and segments are shown in Table~\ref{tab:url_classes}. Segments in curly brackets denote whole URLs that match predefined URL patterns, such as \textsf{GitHub} URLs as denoted by \textit{\{github\}}. These, for instance, are an indicator for available \textit{source code}. Additionally, \textit{documentation} and \textit{artifacts} include all \textit{publications} and \textit{source code} respectively (bold in Tab.~\ref{tab:url_classes}), since we consider publications to be documentation, and source code implicitly constitutes an artifact of software. Even though these heuristics are by no means complete, we tried to cover all cases that we observed by investigating the URLs manually and are convinced to convey a representative round-up of the available resources with this approach.

In addition to the listed datasets, we collected in-links from external websites to the URLs under consideration. These were extracted from the archived German part of the Web under the top-level domain \textit{.de} from 1996 to 2013, which we had full local access to\footnote{This dataset has been provided to us by the Internet Archive in the context of \textit{ALEXANDRIA} (\url{http://alexandria-project.eu})}. Due to scope of this dataset, we performed the in-link analysis only on URLs of domains ending in \textit{.de} as well, assuming that these are better represented in the dataset than arbitrary URLs.

\begin{table}[t!]
  \newcolumntype{C}{>{\centering\arraybackslash}X}
  \newcolumntype{L}{>{\raggedright\arraybackslash}X}
  \newcolumntype{R}{>{\raggedleft\arraybackslash}X}
  \small
  \centering
  
  \caption{Segments extracted from software URLs and grouped into classes.}
  
\begin{tabularx}{\columnwidth}{lL}
\toprule
\textbf{Class}&\textbf{Segments}\\
\midrule
\textbf{source code} & \textit{code, gpl, lgpl, \{R-project\}, \{github\}, \{googlecode\}, \{sourceforge\}, \{cpc\}, \{gpl\}, \{bitbucket\}, \{gnu\}}\\
\textbf{publications} & \textit{publications, papers, journals, publication, article, journal, doi, articles, library, bib, reports, \{acm\}, \{springer\}, \{sciencedirect\}, \{wiley\}, \{cpc\}, \{arxiv\}, \{googlebooks\}, \{ieee\}, \{doi\}, \{manuscriptcentral\}, \{tandfonline\}, \{oxfordjournals\}, \{citeseerx\}}\\
\textbf{updates} & \textit{changelog, history, news, blog}\\
\textbf{documentation} & \textit{doc, documentation, manual, api, reference, handbook, handbuch, referenz, doku, dokumentation, wiki, docs, readme, \textbf{publications}}\\
\textbf{artifacts} & \textit{exe, zip, gz, tar, download, tgz, files, downloads, ftp, \textbf{source code}}\\
\bottomrule
\end{tabularx}
  
  \vspace{-3mm}
  \label{tab:url_classes}
\end{table}

\section{Analysis Results}
\label{sec:results}

The results of our study answer the questions introduced in Section~\ref{sec:problem}. While \textbf{Q1} and \textbf{Q2} focus on software (SW) on the Web in general, \textbf{Q3} and \textbf{Q4} address its coverage by Web archives. In terms of terminology (cp. Sec.~\ref{sec:introduction}), the website of a SW, addressed in the first two questions, generally represents the \textbf{product} or \textbf{product-family}. At the same time, all information on the current state of the website usually refer to the current \textbf{artifact} of the SW, which may or may not be provided itself as download (s. Sec.~\ref{sec:software_on_the_web}). The versions available in a Web archive, addressed by the latter questions, are considered surrogates for the \textbf{artifact} that was prevailing at the time of capture (s. Sec.~\ref{sec:software_in_web_archives}).

\subsection{Software on the Web}
\label{sec:software_on_the_web}

\begin{figure}[t!]
	\centering

	\captionsetup[subfigure]{oneside,margin={0cm,0.5cm}}

	\subfloat[Software references vs. in-links.]{
		\hspace{-13pt}
		\includegraphics[width=0.55\textwidth]{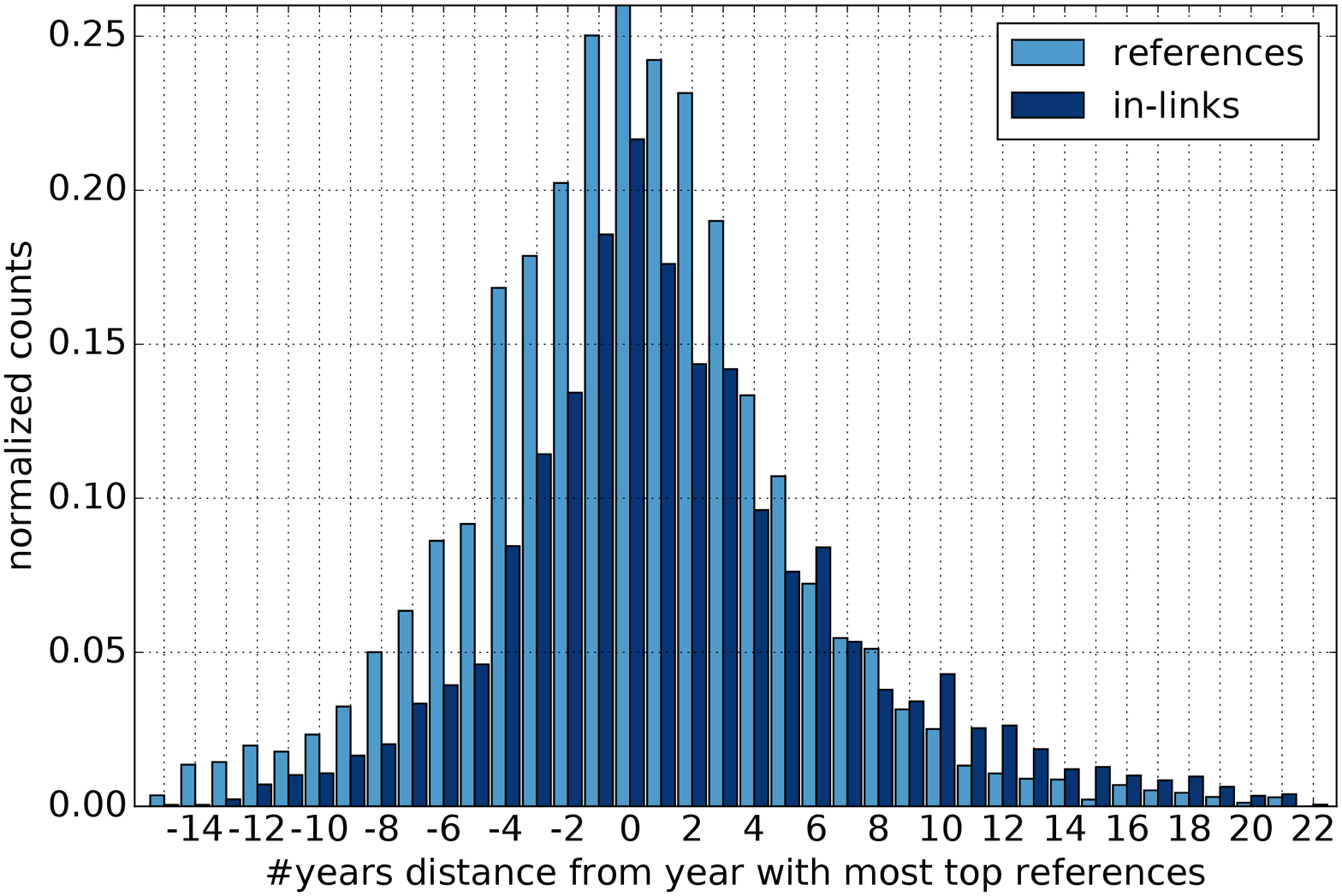}
		\label{fig:inlinks}
	}
	\subfloat[Information on software webpages.]{
		\hspace{-30pt}
		\includegraphics[width=0.55\textwidth]{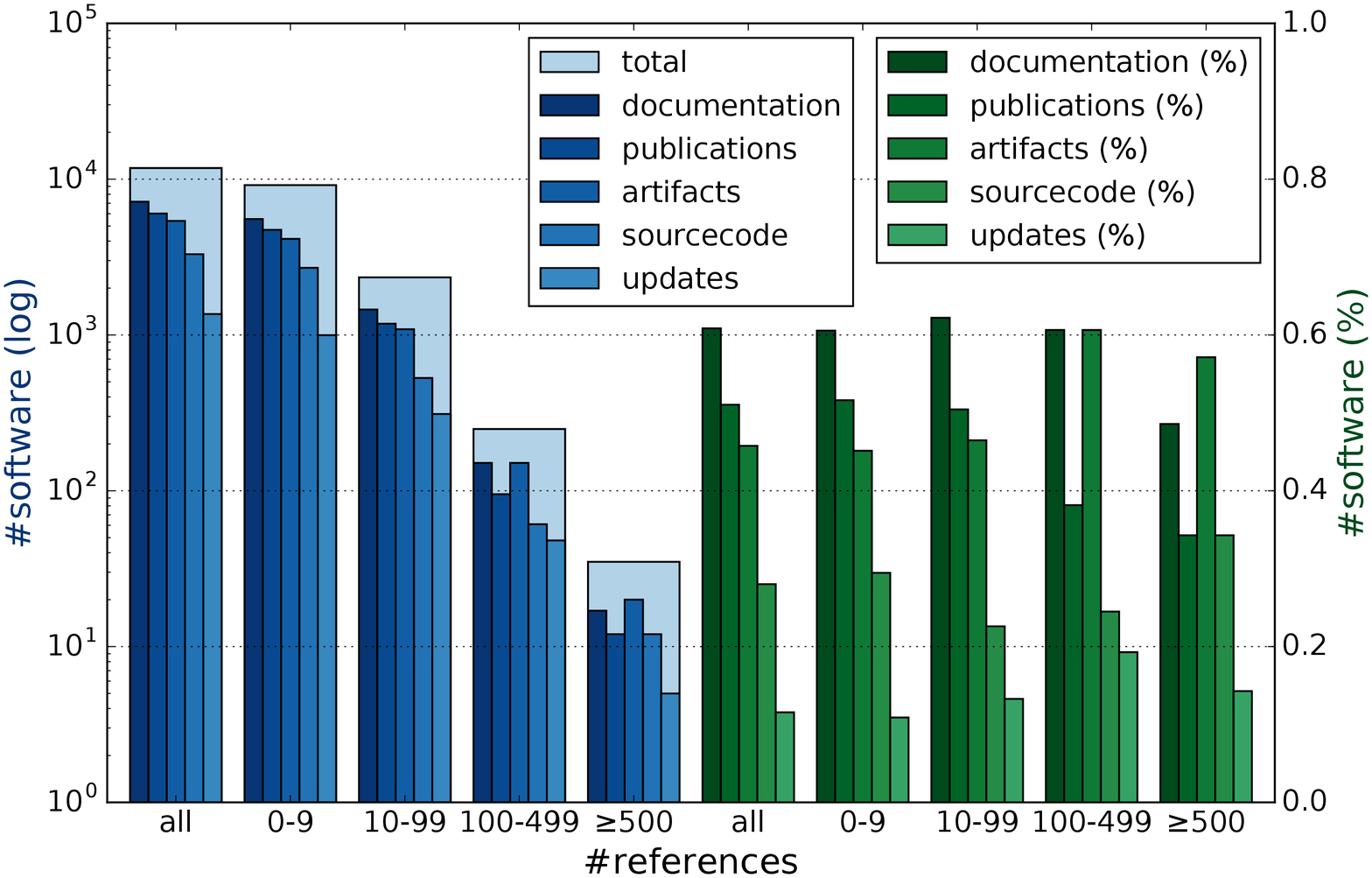}
		\label{fig:websites}
	}

	\caption{Software surrogates on the Web.}
\end{figure}

The first objective of our analysis was to investigate whether the Web reflects real SW at all and how well webpages as potential surrogates represent actual SW (\textbf{Q1}). Our attempt to show such a relation involves the publications referring to a SW over time as well as the in-links to a SW's webpage, which we consider the equivalent to scientific citations on the Web. Figure~\ref{fig:inlinks} illustrates this remarkable correlation, with references slightly ahead of the in-links. The plot has been normalized by the highest number of publications and in-links for a given SW and aligned by the year with most publications for a SW at \textit{x = 0}. It is based on the links extracted from \textit{.de} pages as well as the number of articles in a year among the top 100 publications (cp. Sec~\ref{sec:analysis}). Due to the available link data, only SW with URLs under \textit{.de} was considered.

The fact that the Web lacks slightly behind references in literature suggest that the scientific use of SW leads to visibility and has a strong impact on its popularity. This motives our effort to archive SW's webpages at the time of publication or even the time of use for a publication. Without such a preservation copy, links that were created as a result of a publication may become stale as the SW and its website evolve and the original reference target cannot be recovered. The same applies to SW references in articles, too.

Next, we asked the question of which information can be obtained from the pages (\textbf{Q2}), based on the segments identified in the URLs (cp. Sec~\ref{sec:analysis}). Figure~\ref{fig:websites} shows the total numbers as well as results separated by popularity with respect to the number of publications referencing a SW.

Interesting here is the finding that for around 60\% of the analyzed SW, the corresponding webpage links to some sort of documentation. In many of these cases, publications are available too, or comprise the documentation (cp. Sec.~\ref{sec:analysis}). However, this changes for more popular SW, where a larger fraction of their documentation is independent of publications. Also, as the plot shows, 50\% provide artifacts online, which is again even higher for popular SW. As a result, this trend suggests SW that is well represented on the Web is more prominent and more often used. Therefore, it was surprising to us that we only found source code for around 30\% throughout all popularities. Since only this minority enables a detailed tracing of the development process (s. Sec.~\ref{sec:problem}), it is even more important to store temporal copies of the SW's webpages to get a sense of the development through the available information, such as documentation or update reports. Overall, we can conclude that SW webpages contain valuable information and indeed can serve as surrogates of the actual SW.

\subsection{Software in Web archives}
\label{sec:software_in_web_archives}

\begin{figure}[t!]
	\centering

	\captionsetup[subfigure]{oneside,margin={0cm,0.5cm}}

	\subfloat[Archived software pages.]{
		\hspace{-12pt}
		\includegraphics[width=0.55\textwidth]{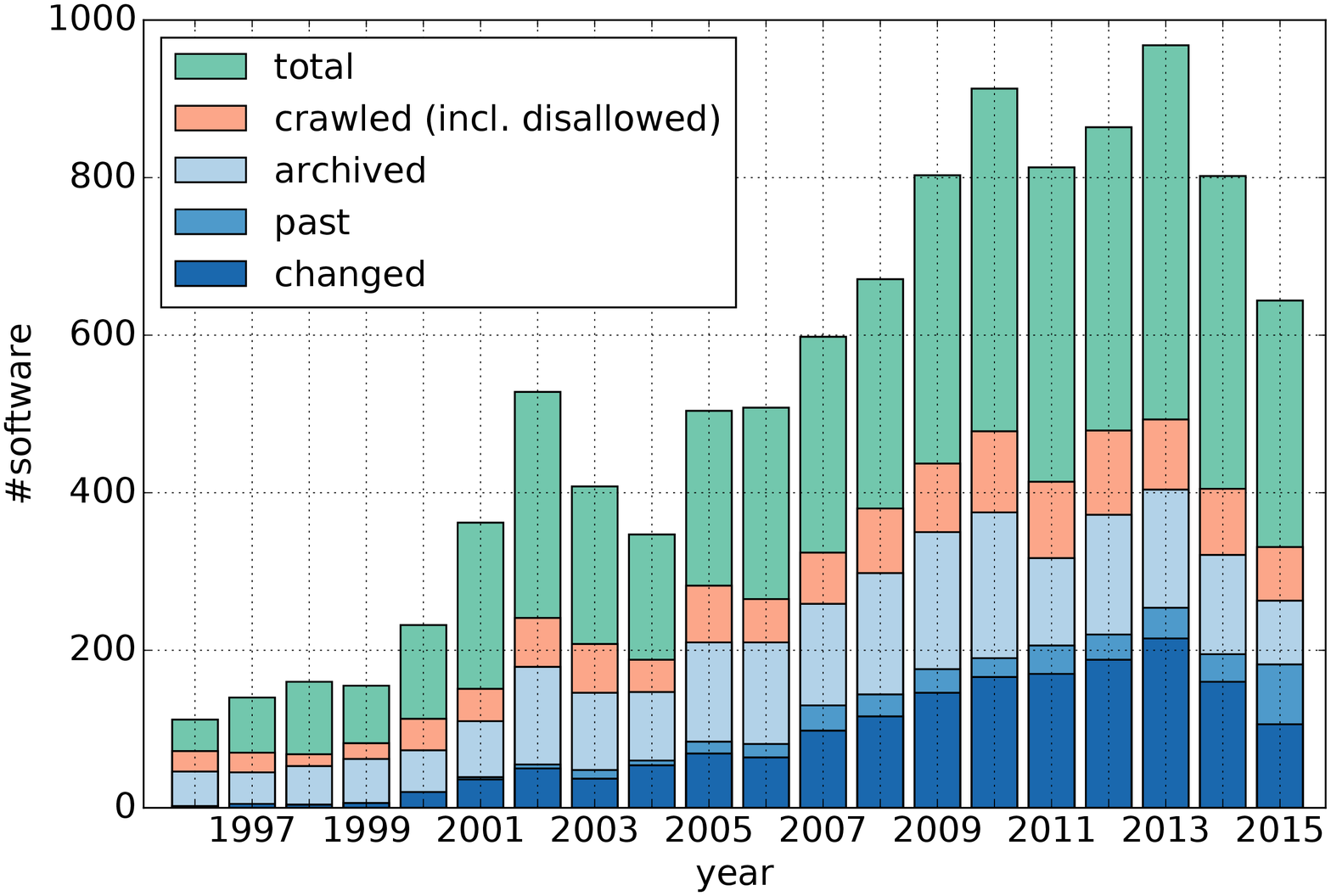}
		\label{fig:archived}
	}
	\subfloat[Temporal archive gap.]{
		\hspace{-25pt}
		\includegraphics[width=0.55\textwidth]{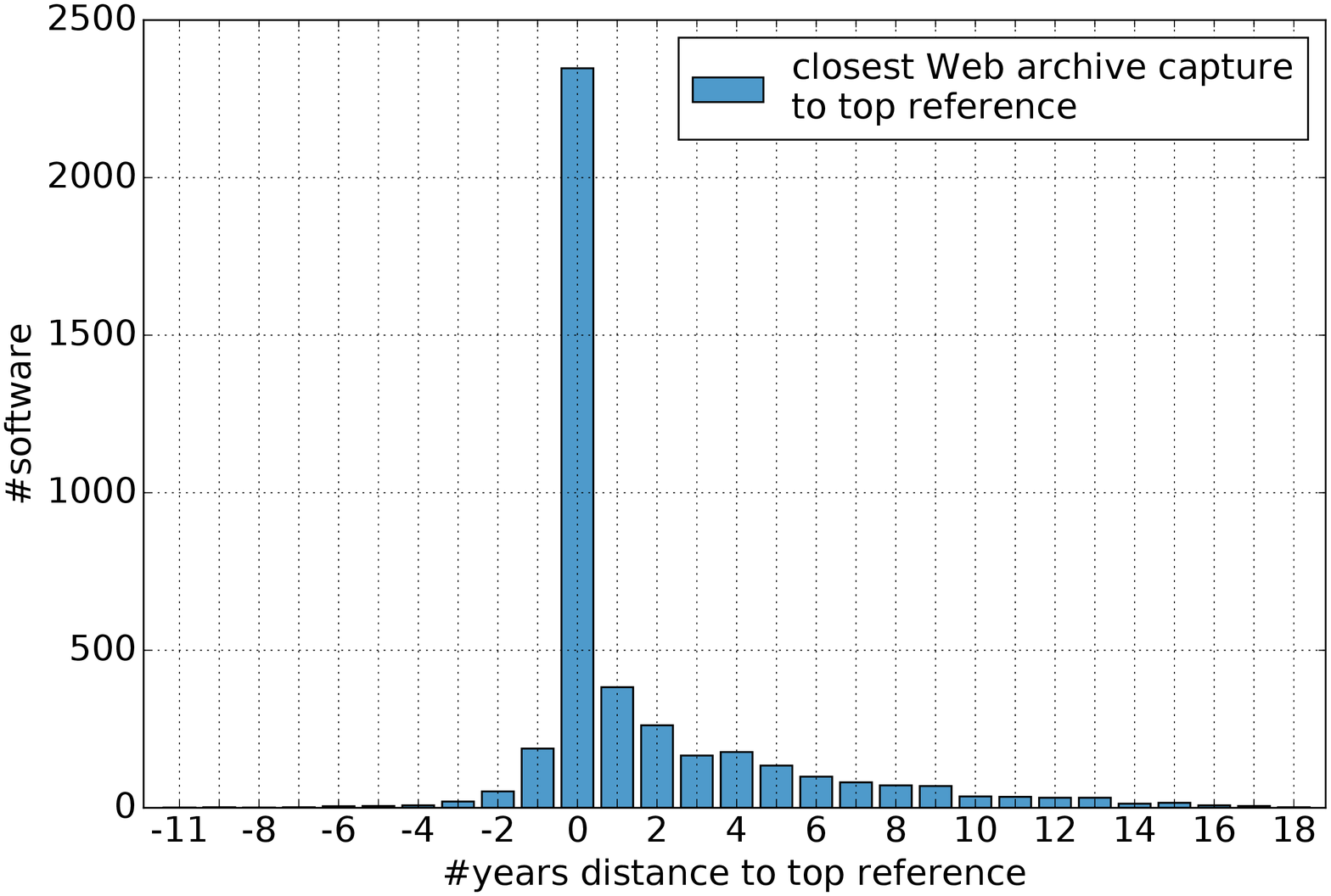}
		\label{fig:gap}
	}

	\caption{Software surrogates in Web archives.}
\end{figure}

For future work, we are planning to develop strategies and mechanisms to create Web archive collections tailored to SW, which cover the above presented information of SW on the Web. However, for SW published and referenced until then, it is valuable to investigate how well existing, generic Web archives have captured SW surrogates on the Web (\textbf{Q3}). The required data for this analysis was obtained from the Internet Archive (s. Sec.~\ref{sec:analysis}). Figure~\ref{fig:archived} shows that almost consistent over time the URLs of around 50\% of the SW under consideration have been captured in their Web archive. 10\% of these were disallowed to be preserved by the \textit{robots.txt} of that website.

While 40\% of preserved contents is still a relatively satisfying number, it gets worse when looking at the fraction of archived captures at the time of the top publication referencing the SW, i.e., the article with most citations (s. \textit{past} in Fig.~\ref{fig:archived}). Although we considered full years, only around 20\% of the SW pages were preserved in that period (\textbf{Q4}). However, due to the efforts at \textsf{swMATH} to replace outdated links (cp. Sec.~\ref{sec:swmath}), this number may actually be higher but not surfaced by our study. At the same time, this would mean that many of the original URLs have been outdated, which in turn suggests a certain development of the corresponding websites. Either way, our findings show the need for a sophisticated Web archiving infrastructure as part of the scientific SW management process, which assigns preserved material on the Web assigned to a specific SW or its reference in a publication, rather than URLs.

Another large portion that is not covered by the number of pages archived in the past are those webpages that were archived shortly before or after the year of the publication. The good news is, this gap is very small as shown in Figure~\ref{fig:gap}. Most pages were captured in the exact year and the remaining were preserved closely around this time with decreasing numbers further away from the year of publication. Thus, by relaxing the time to identify a SW's surrogate in a Web archive to a couple of years around a publication, we can recover even more. Although the representativeness becomes less accurate this way, it can still be helpful to comprehend SW references or reproduce experimental results.

Not exactly surprising, but notable is the fact that almost all pages with archived captures in the past have changed according to the hash/digest of their archived record (s. Fig.~\ref{fig:archived}). This motivates to preserve those copies for future reference.

\section{Conclusion and Outlook}

As shown in our study, the Web reflects software to a remarkable extent, with documentation and artifacts on a considerable number of webpages. Hence, the Web can indeed serve as surrogate of actual software. However, only for about a half of the analyzed mathematical software, the corresponding webpages are preserved by an existing Web archive. To fix this in the future, we propose establishing new infrastructures to actively archive software surrogates on the Web at the time of use. The author of an article should be able to do this on demand and be provided with a handle to reliably reference the archived surrogate. As a first step, Web archives may be integrated with software directories, such as \textsf{swMATH}, to link software in existing scientific articles based on their publication date. Eventually, explicit references will improve the management of software in scientific publications.

\bibliographystyle{IEEEtranN}
\bibliography{references}

\end{document}